\documentclass[runningheads]{llncs}
\usepackage[T1]{fontenc}
\usepackage{graphicx}
\usepackage{amsmath}
\usepackage{amssymb}
\usepackage{booktabs}
\usepackage{algorithm}
\usepackage{algpseudocode}
\usepackage{multirow}
\usepackage{csquotes}
\usepackage[normalem]{ulem}
\usepackage{cite}
\useunder{\uline}{\ul}{}
\usepackage[pagebackref,breaklinks,colorlinks]{hyperref}
\begin{document}
\title{F2FLDM: Latent Diffusion Models with Histopathology Pre-Trained Embeddings for Unpaired Frozen Section to FFPE Translation}
\titlerunning{F2FLDM}
%
\author{Man M. Ho$^{1}$ \qquad 
Shikha Dubey $^{1}$ \qquad 
Yosep Chong$^{2,3}$ \\
Beatrice Knudsen$^{3,4}$ \qquad 
Tolga Tasdizen$^{1,5}$}
\authorrunning{Ho et al.}
%
\institute{
$^{1}$ Scientific Computing and Imaging Institute, University of Utah, USA \\
$^{2}$ The Catholic University of Korea College of Medicine, Korea \\
$^{3}$ Departmant of Pathology, University of Utah, USA \\
$^{4}$ Huntsman Cancer Institute, University of Utah Health, USA \\
$^{5}$ Department of Electrical and Computer Engineering, University of Utah, USA
}
\maketitle              
\begin{abstract}
The Frozen Section (FS) technique is a rapid and efficient method, taking only 15-30 minutes to prepare slides for pathologists' evaluation during surgery, enabling immediate decisions on further surgical interventions. However, FS process often introduces artifacts and distortions like folds and ice-crystal effects. In contrast, these artifacts and distortions are absent in the higher-quality formalin-fixed paraffin-embedded (FFPE) slides, which require 2-3 days to prepare. While Generative Adversarial Network (GAN)-based methods have been used to translate FS to FFPE images (F2F), they may leave morphological inaccuracies with remaining FS artifacts or introduce new artifacts, reducing the quality of these translations for clinical assessments. In this study, we benchmark recent generative models, focusing on GANs and Latent Diffusion Models (LDMs), to overcome these limitations. We introduce a novel approach that combines LDMs with Histopathology Pre-Trained Embeddings to enhance restoration of FS images. Our framework leverages LDMs conditioned by both text and pre-trained embeddings to learn meaningful features of FS and FFPE histopathology images. Through diffusion and denoising techniques, our approach not only preserves essential diagnostic attributes like color staining and tissue morphology but also proposes an embedding translation mechanism to better predict the targeted FFPE representation of input FS images. As a result, this work achieves a significant improvement in classification performance, with the Area Under the Curve rising from 81.99\% to 94.64\%, accompanied by an advantageous CaseFD. This work establishes a new benchmark for FS to FFPE image translation quality, promising enhanced reliability and accuracy in histopathology FS image analysis. Our work is available at \url{https://minhmanho.github.io/f2f\_ldm/}.
\end{abstract}
\section{Introduction}

\begin{figure*}[t]
    \centering
    \includegraphics[width=\textwidth]{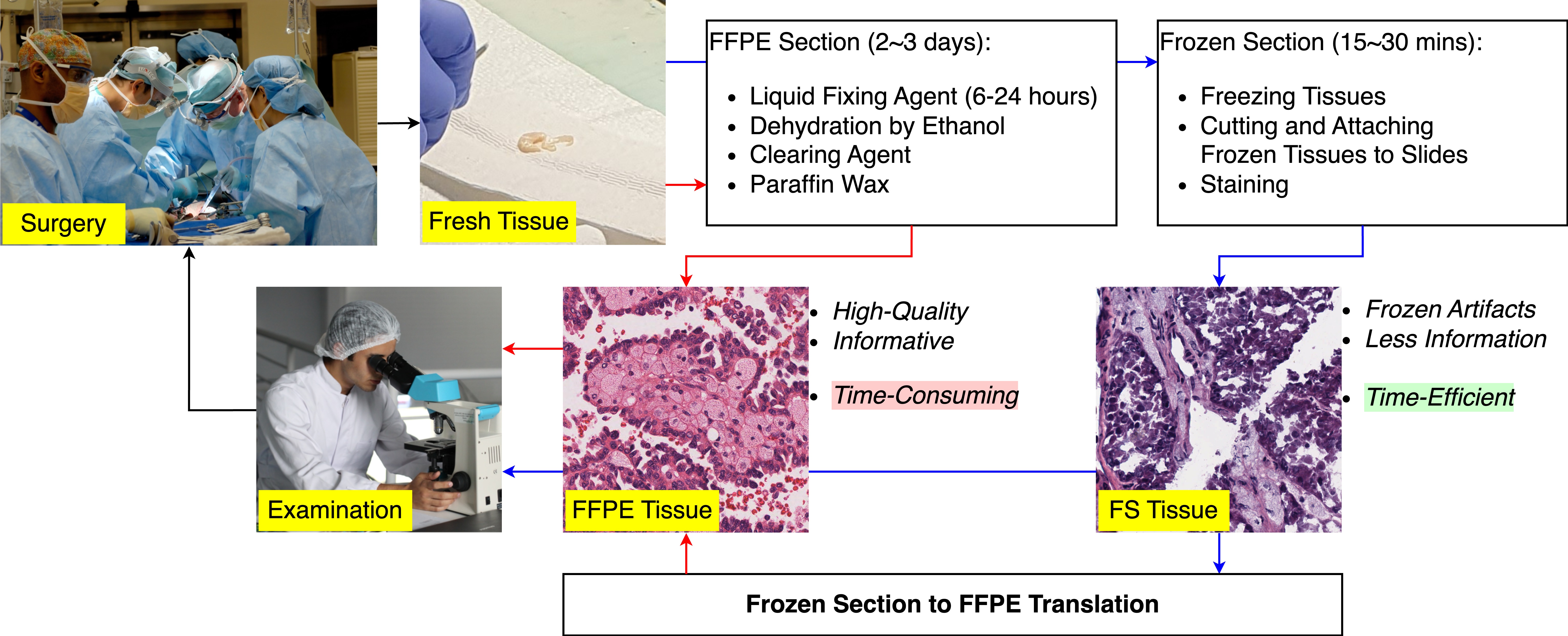}
    \caption{Overview of FS and FFPE processes and our motivation. }
    \label{fig:motiv}
\end{figure*}

Histopathology, the microscopic examination of tissues to diagnose diseases, relies heavily on the Formalin-Fixed Paraffin-Embedded (FFPE) technique for producing high-quality tissue slides. Considered the gold standard, FFPE offers detailed, artifact-free slides crucial for accurate diagnosis. Yet, the FFPE process is notably slow, often requiring two to three days to prepare slides, making it unsuitable for surgeries that necessitate immediate decisions. Therefore, Frozen Section (FS) procedure is preferred due to its quick processing time. FS allows pathologists to examine tissue samples within minutes, providing surgeons with instant information that can influence surgical decisions and potentially improve patient outcomes. Despite its speed, FS image quality suffers from the introduction of artifacts such as tissue folds and ice crystals. These can obscure crucial histological details and complicate diagnoses, as described in Figure \ref{fig:motiv}.

\begin{figure*}[t]
    \centering
    \includegraphics[width=\textwidth]{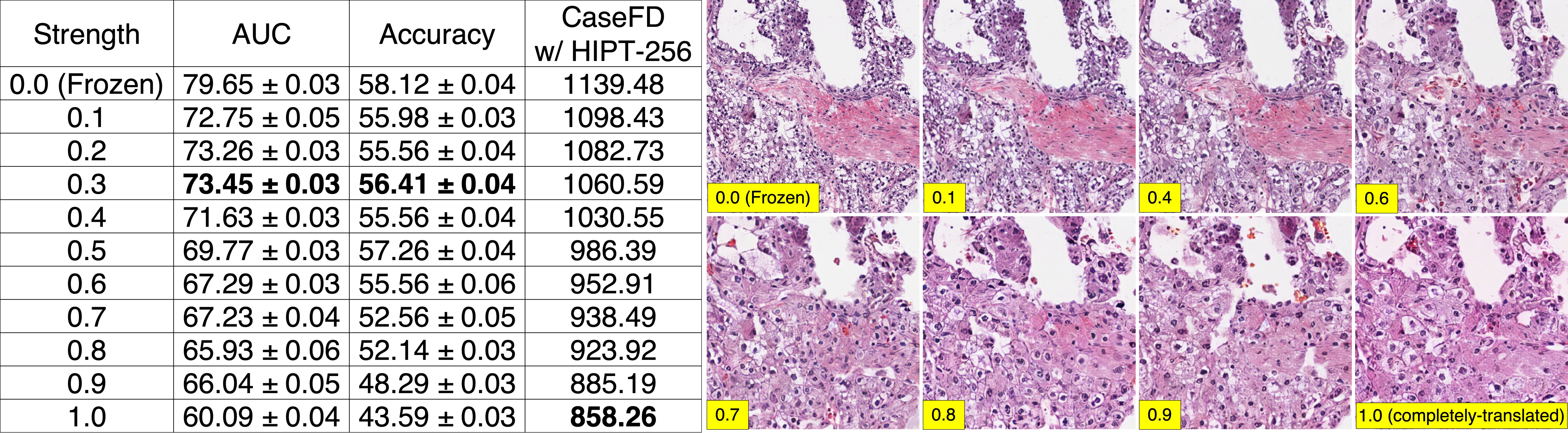}
    \caption{Performance on CycleDiffusion-restored slides on kidney subtype classification, including Area Under the Curve (AUC) and accuracy, alongside case-wise Fréchet Distance in the HIPT-256 feature space (FD-HIPT256). The higher \textbf{strength}, the more added noise and denoising timesteps, the closer to FFPE domain.}
    \label{fig:cd_issue}
\end{figure*}

The integration of Artificial Intelligence (AI) into histological analysis has been revolutionized by Generative Adversarial Networks (GANs) \cite{isola2017image,CycleGAN2017,park2020contrastive,torbunov2023uvcgan,torbunov2023uvcgan2,aiffpe,kang2023gan4,sc-gan} and Latent Diffusion Models (LDMs) \cite{ho2020denoising,rombach2022high,wu2022cyclediffusion,moghadam2023morphology}, facilitating domain-to-domain image translation without paired samples. These technologies have made significant strides, yet occasionally struggle to maintain the detailed accuracy crucial for disease diagnosis in histological images. Challenges include not fully removing artifacts from Frozen Sections (FS) or introducing new artifacts. Specifically, the approach described in \cite{aiffpe} aims to improve the quality of FS images towards FFPE standards by employing Unpaired Contrastive Translation. This method is primarily focused on correcting artifacts using an attention mechanism and self-regulation to preserve clinically relevant features. Nonetheless, it does not entirely overcome the challenge of maintaining the complex tissue structure characteristic of FFPE images, which is essential for accurate diagnoses. Moreover, diffusion models employing a Gaussian latent space formulation \cite{wu2022cyclediffusion} present a novel approach to image translation. By unifying the latent space and exploiting its cyclic nature, this method facilitates image translation across different domains while preserving content and adjusting domain-specific features. While this process improves Fréchet Distance (FD) scores, it decreases the accuracy of downstream tasks, such as classification, as shown in Figure \ref{fig:cd_issue}.


Given these challenges, our study introduces a novel framework for unpaired FS to FFPE image translation, leveraging the capabilities of LDMs enhanced with Histopathology Pre-Trained Embeddings. This approach aims to produce high-fidelity FS and FFPE images by leveraging text descriptions and pre-trained embeddings. To address the challenge of unpaired translation - where the target FFPE embeddings are absent, we employ a GAN-based U-style Fully Connected Network. This network effectively converts FS embeddings into their FFPE versions, offering improved guidance for generating authentic FFPE images. This novel strategy promises to significantly improve the translation and artifact restoration in FS images, addressing common issues such as FS artifact presence and morphological inaccuracies, thereby improving the accuracy and reliability of histological  analysis for clinical assessments.

Our contributions are as follows: 
1) Benchmarking latest generative models to address the FS to FFPE image translation challenge.
2) Developing a FS to FFPE image translation framework that improves artifact restoration in FS images, utilizing LDMs with Histopathology Pre-Trained Embeddings, including a mechanism for FS to FFPE embedding translation to overcome the absence of direct FFPE embeddings.
3) Establishing robust evaluation metrics for FS to FFPE image translation , utilizing case-wise Fréchet Distance (CaseFD) within a histopathology pre-trained latent space, alongside downstream classification tasks, evaluating how translation enhances classification accuracy.
4) Significantly outperforming existing state-of-the-art solutions, such as AI-FFPE, UVCGAN2, and CycleDiffusion in downstream classification performance with favorable CaseFD.

\section{Methods}
\begin{figure*}[t]
    \centering
    \includegraphics[width=\textwidth]{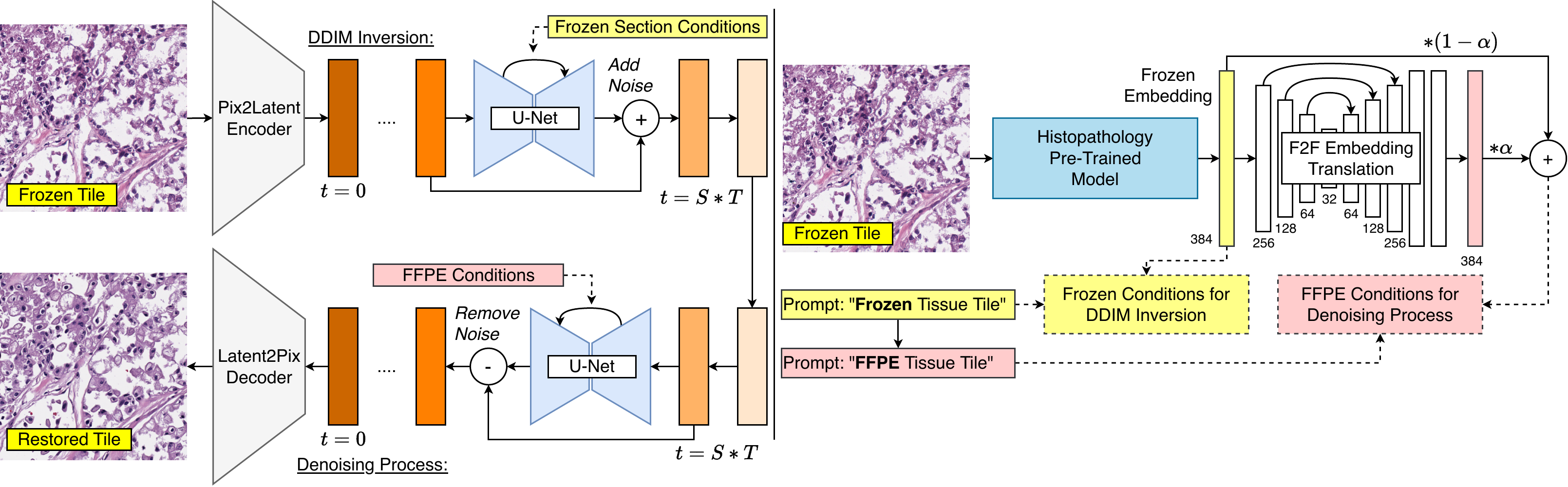}
    \caption{Overview of our FS to FFPE image translation framework.}
    \label{fig:overview}
\end{figure*}
In this study, we present a novel FS to FFPE image translation framework that leverages Latent Diffusion Models (LDMs) and Histopathology Pre-Trained Embeddings. By conditioning LDMs with text descriptions and embeddings, our denoising model effectively captures and translates the unique characteristics of FS and FFPE images with high fidelity. The process begins with transforming FS images into noisy versions via DDIM inversion, tailored with FS-specific data. A denoising U-Net then restores these images, which are further refined into FFPE equivalents using FFPE-specific data and a GAN-based network for embedding translation, as shown in Figure \ref{fig:overview}. This method not only maximizes the LDMs' potential in producing authentic FFPE images but also focuses on artifact restoration, crucial for improving histopathology image analysis.

\subsection{Latent Diffusion Models with Histopathology Pre-Trained Embeddings}

Our research utilizes Hematoxylin and Eosin (H\&E) stained images $x \in \mathbb{R}^{H \times W \times 3}$, applying Latent Diffusion Models (LDMs) enhanced with histopathology pre-trained embeddings for accurate FS to FFPE image translation. FS images ($x_{\text{fs}}$) and Formalin-Fixed Paraffin-Embedded (FFPE) images ($x_{\text{ffpe}}$) are associated with respective text descriptions $p$ and embeddings $e = \mathcal{V}(x)$, where $\mathcal{V}$ denotes models pre-trained on histopathology data, facilitating precise feature capture and translation. Through a pix2latent encoder ($\mathcal{E}$), images are converted into simpler latent representations ($z$), which are then refined back into detailed images by a latent2pix decoder ($\mathcal{D}$), as described in Figure \ref{fig:overview}.

The translation process introduces Gaussian noise to these latent representations, incrementally adjusted over $T$ timesteps. A specialized denoising U-Net ($\epsilon_{\theta}$), trained to predict and remove the added noise guided by $p$ and $e$, then accurately restores FS/FFPE images. The training loss function for the denoising U-Net is simplified as: $\mathcal{L}_{LDM} = \mathbb{E}_{z,\epsilon \sim \mathcal{N}(0, I),t,p,e}\left[||\epsilon-\epsilon_{\theta}(z_t, t, p, e)||^2_2\right]$, where $z_t$ denotes the noisy version of $z_0$ at the timestep $t$, sampled uniformly from the set $\{1,...,T\}$, where $T=1000$. To ensure efficient training and effective FS/FFPE image generation, we adopt the pre-trained Stable Diffusion XL (SDXL) \cite{podell2023sdxl}, incorporating extra linear layers for processing pre-trained embeddings and fine-tuning their text encoders and denoising U-Net with Low-Rank Adaptation (LoRA) \cite{hu2021lora}.

\subsection{Frozen Section to FFPE Image Translation}

Leveraging the refined SDXL framework, we developed a FS to FFPE image translation method, $\text{F2F}: x_{\text{fs}} \rightarrow x_{\text{ffpe}}$, using histopathology-specific text descriptions $p$ and embeddings $e$. The process starts with encoding FS images into latent representations, $z_{\text{fs\_0}} = \mathcal{E}(x_{\text{fs}})$, which are then noised by $\epsilon_{\theta}(z_t, t, p_{\text{fs}}, e_{\text{fs}})$ via DDIM inversion with a strength $S \in [0.0, 1.0]$, $T=50$, and $t = 0, 1, 2, \ldots, S*T$. Next, a U-style fully connected network $G$, trained with WGAN-GP \cite{gulrajani2017wgan} in cycle fashion \cite{CycleGAN2017}, translates FS embeddings $e_{\text{fs}}$ to FFPE equivalents $\hat{e}_{\text{ffpe}}$, blending them using an interpolation weight $\alpha$ to maintain FS image identity. The final step involves progressively eliminating the added noise $\epsilon_{\theta}(z_t, t, p_{\text{ffpe}}, \hat{e}_{\text{ffpe}})$ from $z_{\text{S*T}}$ to $t=0$, culminating in the latent base $\hat{z}_{\text{ffpe\_0}}$, which is then transformed into the FFPE image $\hat{x}_{\text{ffpe}} = \mathcal{D}(\hat{z}_{\text{ffpe\_0}})$, as shown in Figure \ref{fig:overview}. 


\begin{table}[t]
\centering
\caption{Our train, validation, and test sets with three kidney subtypes equally distributed over cases.}
\label{supp_tab:dataset}
\resizebox{0.8\textwidth}{!}{%
\begin{tabular}{|cc|cc|cc|}
\hline
\multicolumn{2}{|c|}{Train} & \multicolumn{2}{c|}{Validation} & \multicolumn{2}{c|}{Test} \\ \hline
\multicolumn{1}{|c|}{\#cases} & \#patches (1024x1024) & \multicolumn{1}{c|}{\#cases} & \#patches (1024x1024) & \multicolumn{1}{c|}{\#cases} & \#patches (4096x4096) \\ \hline
\multicolumn{1}{|c|}{180} & 314,114 & \multicolumn{1}{c|}{39} & 70,484 & \multicolumn{1}{c|}{39} & 636 \\ \hline
\end{tabular}%
}
\end{table}

\textbf{Enhancing Translation with Classifier-Free Guidance (CFG) and L0 Regularization}. 
Our exploration into LDMs' capacity for artifact removal in FS images, while maintaining and refining morphology, led us to innovate in guiding the translation process. Increasing the Guidance Scale (GS) enhances the FFPE feature presence and minimize FD between the translated and real FFPE image distributions in latent space. However, an unmoderated increase in GS risks distorting FS image morphology, potentially reducing downstream classification performance, as discussed in Figure \ref{fig:cd_issue}. Furthermore, reliance on unconditional predicted noise in CFG \cite{ho2022cfg} could introduce artifacts. In pursuit of a balanced translation that respects both FFPE conditions and the inherent creativity of the model, we replace the unconditional noise with noise conditioned by FFPE-translated embeddings and apply L0 Regularization \cite{han2024proxedit} to harmonize the conditional noise with embedding-guided noise as follows:

\begin{equation}
\label{eq_reg}
   \hat{\epsilon}_t = \epsilon_{\theta}(z_t, t, \varnothing, \hat{e}_{\text{ffpe}}) + \text{GS}* \text{prox}_{\lambda}\big(\epsilon_{\theta}(z_t, t, p_{\text{ffpe}},\hat{e}_{\text{ffpe}}) - \epsilon_{\theta}(z_t, t, \varnothing, \hat{e}_{\text{ffpe}})\big)
\end{equation}

where $\text{prox}_{\lambda}(d) = d$ if $|d| > \sqrt{2\lambda}$, and $0$ otherwise. $\lambda$ is set as the 70\% quantiles of the absolute values of the noise difference, in line with \cite{han2024proxedit}. Note that our model without $\text{prox}_{\lambda}$ will apply all noise changes.

\section{Experiments}

\textbf{Datasets}. We use the TCGA-Kidney dataset, including 258 cases evenly spread across three kidney subtypes (ccRCC, ChRCC, PRCC) with 516 slides in total. The division is 180 cases for training (70\%), 39 for validation (15\%), and 39 for testing (15\%) (details described in Table \ref{supp_tab:dataset}). Training and validation use $1024\times1024$ patches, while $4096\times4096$ patches are for slide-level test evaluation \cite{chen2022scaling}. Notably, our evaluation focuses on restoring one-by-one $1024\times1024$ patches from these larger test set patches, and only approximate $200$ test set patches are utilized for efficiently evaluating our ablation models.

\begin{figure*}[t]
    \centering
    \includegraphics[width=\textwidth]{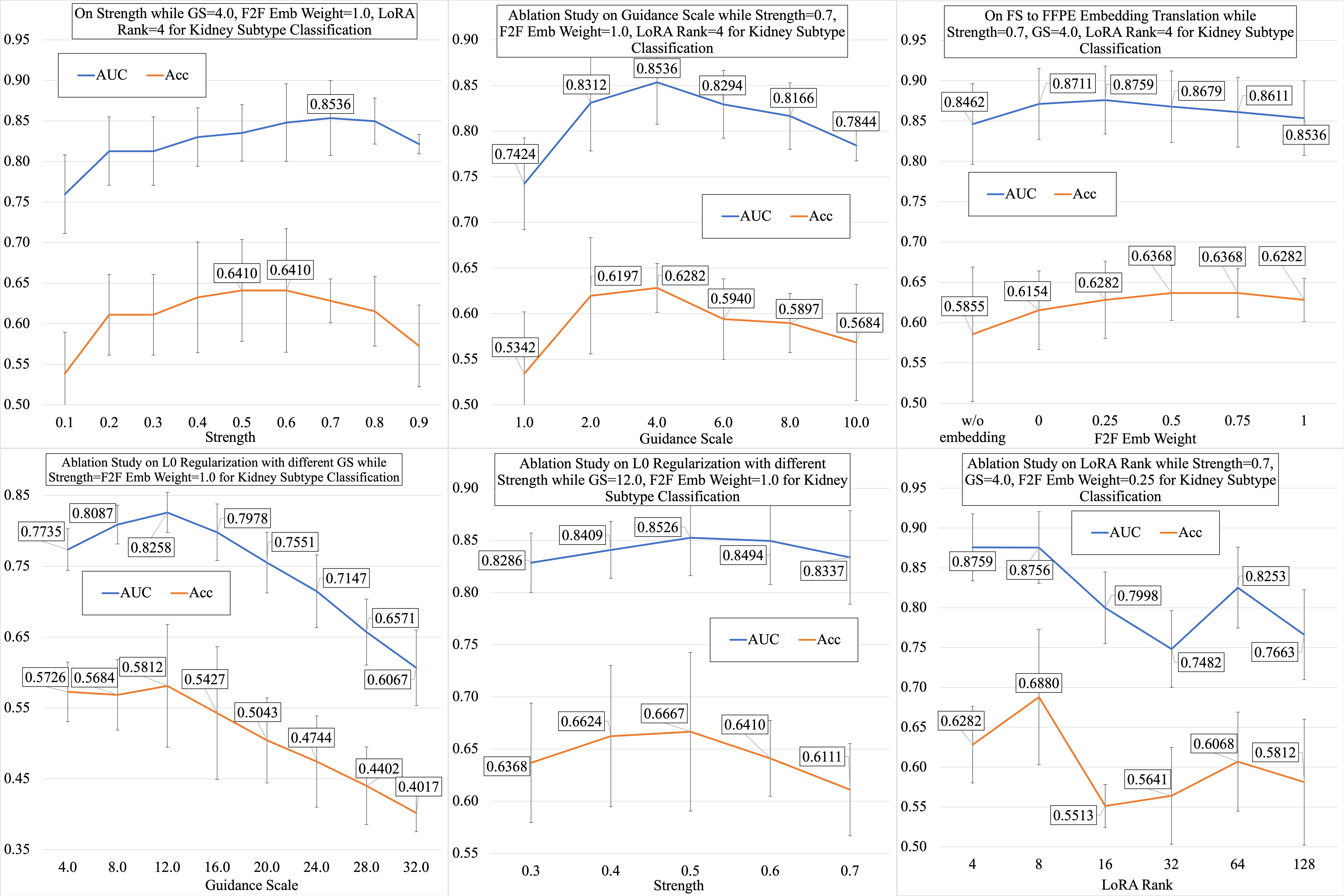}
    \caption{Ablation Studies on classifier-free Guidance Scale (GS), Strength, FS to FFPE Embedding Translation, LoRA rank, and L0 Regularization in Restoration of Artifacts in FS images, evaluated on downstream kidney subtype classfication in macro-averaged Area Under the Curve (AUC) and sample-wise Accuracy (Acc).}
    \label{fig:chart_all}
\end{figure*}

\textbf{Training Details}. The FS to FFPE image translation is fine-tuned using the AdamW optimizer with a learning rate of $0.0001$, $\beta_1 = 0.9$, $\beta_2 = 0.999$, and a weight decay of $0.01$. Training is conducted with a batch size of $1$ on $512\times512$ images. Ablation models are fine-tuned over 50,000 iterations (approximately 7 hours), while the final models including ours and previous works are refined across 150,000 iterations (21 hours) leveraging NVIDIA RTX A6000 GPUs.

\textbf{Evaluation Metrics}. While distribution-based metrics like the Fréchet Distance (FD) are standard for assessing generated images, they do not account for the morphological accuracy crucial for clinical evaluations. To address this, we introduce the Case-wise Fréchet Distance (CaseFD) to compute FD between translated and real images within the same case in the latent space by pre-trained models such as DINOv2 ViT-L14 \cite{stein2024exposing}, HIPT-256 \cite{chen2022scaling}, and ViT-DINO \cite{kang2022benchmarking}. These models are pre-trained on general images, TCGA FFPE slides, and both TCGA and TULIP datasets, respectively. Moreover, we implement a Fully-Connected Multiple Instance Learning (MIL) network based on the HIPT \cite{chen2022scaling}, specifically for kidney subtyping on FFPE slides, using a 6-fold leave-one-out cross-validation from our dataset. This approach assesses whether FS to FFPE image translation enhances kidney subtype classification accuracy, providing an in-depth evaluation of its clinical relevance.

\begin{figure*}[t]
    \centering
    \includegraphics[width=\textwidth]{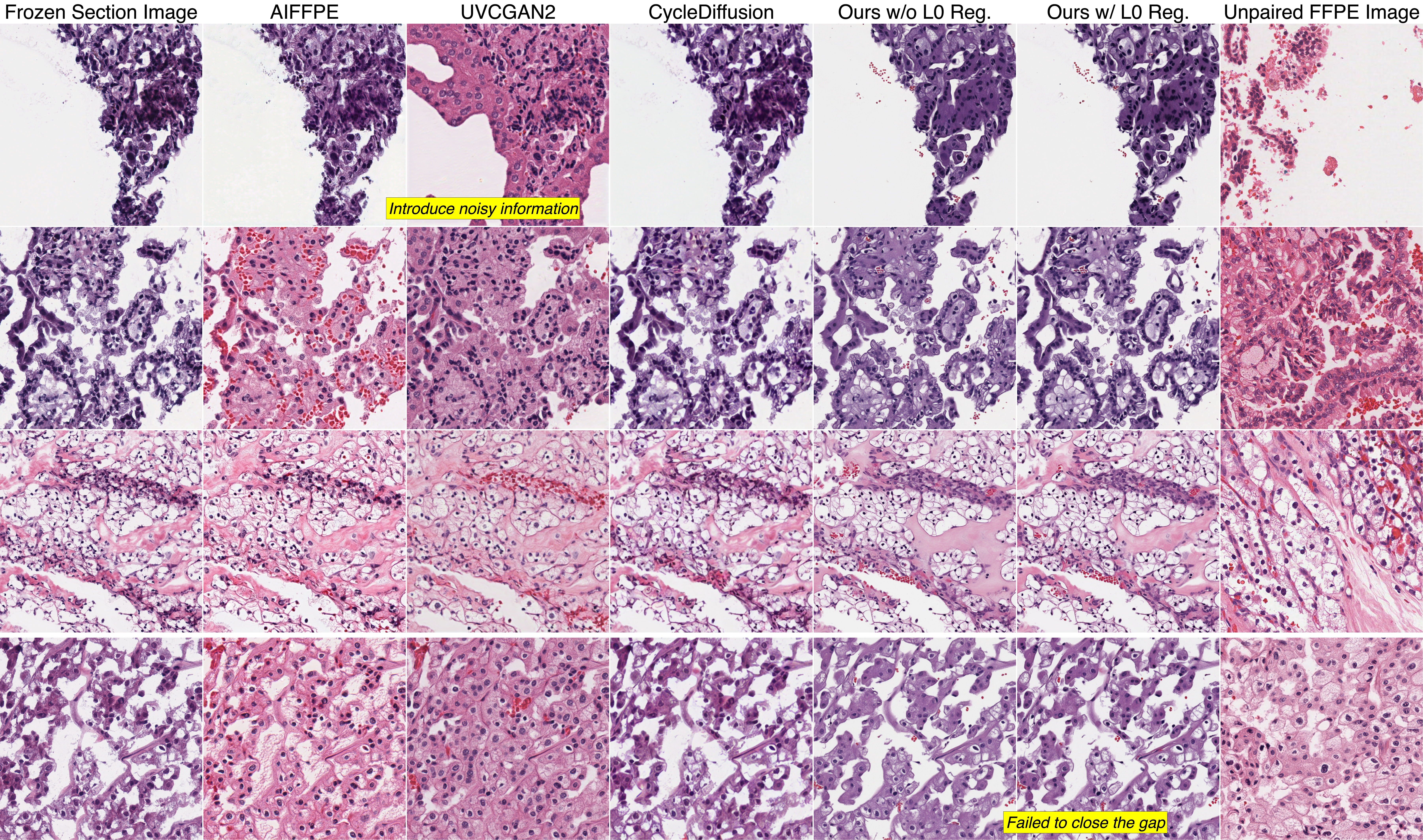}
    \caption{A qualitative comparison between AIFFPE \cite{aiffpe}, UVCGAN2 \cite{torbunov2023uvcgan2}, CycleDiffusion \cite{wu2022cyclediffusion}, and ours. Unpaired FFPE images are from a different tissue.}
    \label{fig:qual_comp}
\end{figure*}

\textbf{On Strength, Guidance Scale (GS), and LoRA Rank}. Tuning these hyperparameters is crucial for optimal translation. As shown in the top-left, top-middle, and bottom-right charts of Figure \ref{fig:chart_all}, adjusting Strength and GS demonstrates a bell-shaped relationship with AUC and accuracy performance, highlighting their impact on downstream classification tasks. These adjustments reveal a trade-off: while increasing Strength and GS can enhance the FFPE-like appearance of images (see Figure \ref{fig:cd_issue}), it can also lead to the introduction of new artifacts, negatively affecting performance. A higher LoRA rank enhances domain adaptation but requires more training time, affecting efficiency. After considering both AUC and accuracy, we identify the optimal settings for our final models as $S=0.7$, $GS=4.0$, and LoRA rank of 8. A qualitative result can be found in Supplementary Document.

\textbf{On Histopathology Pre-Trained Embeddings and Translation}. Integrating histopathology pre-trained embeddings improves FS to FFPE image translation, boosting AUC to $0.8711$ from $0.8462$. However, FS embeddings alone do not fully match FFPE image characteristics. To address this, we introduced a translation mechanism using U-style fully-connected layers based on \cite{CycleGAN2017,gulrajani2017wgan}, predicting FFPE embeddings while maintaining FS identity through an interpolation weight $\alpha$. Choosing $\alpha$ based on AUC and accuracy, we found that a $\alpha=0.25$ further raised AUC to $0.8759$ and accuracy to $0.6282$. See Supplementary Document for qualitative outcomes.

\textbf{On L0 Regularization}. To mitigate new artifacts from increasing Strength and GS, we adopt L0 Regularization on noise difference between conditional noise and embedding-guided noise for image generation guidance, as detailed in Equation \ref{eq_reg}. This ensures significant FFPE-conditioned changes align with the latent representation, disregarding minor and noisy alterations. Consequently, the translation is more robust, resulting lower $S=0.5$ - faster translation and higher $GS=12.0$ - more robust FFPE patterns, as shown in the bottom-left and middle-left charts of Figure \ref{fig:chart_all}. Refer to Supplementary Document for a qualitative result.

\begin{table}[t]
\caption{A quantitative comparison between AIFFPE \cite{aiffpe}, UVCGAN2 \cite{torbunov2023uvcgan2} and CycleDiffusion \cite{wu2022cyclediffusion}, and ours. This work outperforms previous works on the downstream kidney subtype classfication in macro-averaged Area Under the Curve (AUC) and sample-wise Accuracy (Acc), while obtaining the favorable Case-wise Fréchet Distances (CaseFD). \textbf{Bold}/{\ul underlined} values indicate \textbf{best}/{\ul second-best} performance.}
\label{tab:quan_comp}
\resizebox{\textwidth}{!}{%
\begin{tabular}{lccccccc}
\cline{1-1} \cline{3-4} \cline{6-8}
\multicolumn{1}{c}{Method} &  & AUC & Acc &  & \begin{tabular}[c]{@{}c@{}}CaseFD w/ \\ DINOv2 ViT-L14 \cite{stein2024exposing} \end{tabular} & \begin{tabular}[c]{@{}c@{}}CaseFD w/ \\ HIPT-256 \cite{chen2022scaling} \end{tabular} & \begin{tabular}[c]{@{}c@{}}CaseFD w/ \\ ViT-DINO \cite{kang2022benchmarking} \end{tabular} \\ \cline{1-1} \cline{3-4} \cline{6-8} 
FFPE &  & 94.63 ± 0.02 & 88.89 ± 0.03 &  & $\infty$ & $\infty$ & $\infty$ \\
Frozen Section &  & 81.99 ± 0.03 & 61.97 ± 0.08 &  & 546.86 & 1044.24 & 1581.22 \\ \cline{1-1} \cline{3-4} \cline{6-8} 
AIFFPE \cite{aiffpe} &  & 75.46 ± 0.04 & 62.82 ± 0.03 &  & 554.43 & 887.47 & 1243.67 \\
UVCGAN2 \cite{torbunov2023uvcgan2} &  & 84.89 ± 0.01 & 70.09 ± 0.03 &  & \textbf{513.34} & \textbf{808.47} & \textbf{1205.63} \\
CycleDiffusion \cite{wu2022cyclediffusion} &  & 70.55 ± 0.02 & 53.42 ± 0.05 &  & 621.69 & 972.96 & 1486.58 \\
Ours w/o L0 Reg. &  & \textbf{94.64 ± 0.01} & {\ul 73.5 ± 0.06} &  & {\ul 544.85} & 839.67 & {\ul 1235.65} \\
Ours w/ L0 Reg. &  & {\ul 94.26 ± 0.03} & \textbf{80.34 ± 0.07} &  & 546.65 & {\ul 822.66} & 1240.07 \\ \cline{1-1} \cline{3-4} \cline{6-8} 
\end{tabular}%
}
\end{table}

\textbf{Qualitative and quantitative results}. 
Following our ablation studies, we subsequently train our model until 150,000 iterations with a LoRA rank of 8, applying the translation to all test patches with and without L0 Regularization for comprehensive evaluation. We conduct a comparative analysis against AIFFPE \cite{aiffpe}, UVCGAN2 \cite{torbunov2023uvcgan2}, and CycleDiffusion \cite{wu2022cyclediffusion}, across qualitative results, CaseFB, and downstream kidney classification accuracy (exclusively trained on FFPE images). Qualitatively, UVCGAN2 showed a better performance in mimicking FFPE image characteristics, notably in filling white gaps with tissue textures. However, this was accompanied by the introduction of additional artifacts that reduced its effectiveness for clinical assessments. Conversely, our models substantially improved histological  details without significant artifact introduction. Quantitatively, while UVCGAN2 led in CaseFD (\textbf{513.34}, \textbf{808.47}, and \textbf{1205.63}) and demonstrated improvements of \textbf{+2.9} in AUC and \textbf{+8.3} in accuracy, our model with L0 Regularization outperformed in enhancing AUC by \textbf{+12.27} and accuracy by \textbf{+18.55}, achieving favorable CaseFD results (\textbf{546.65}, \textbf{822.66}, and \textbf{1240.07}), as described in Table \ref{tab:quan_comp}. More results and a comparison to CycleDiffusion with similar hyper-parameters are in Supplementary Document and Figure \ref{fig:supp_cd_comp}, respectively.

\begin{figure}[t]
    \centering
    \includegraphics[width=0.9\textwidth]{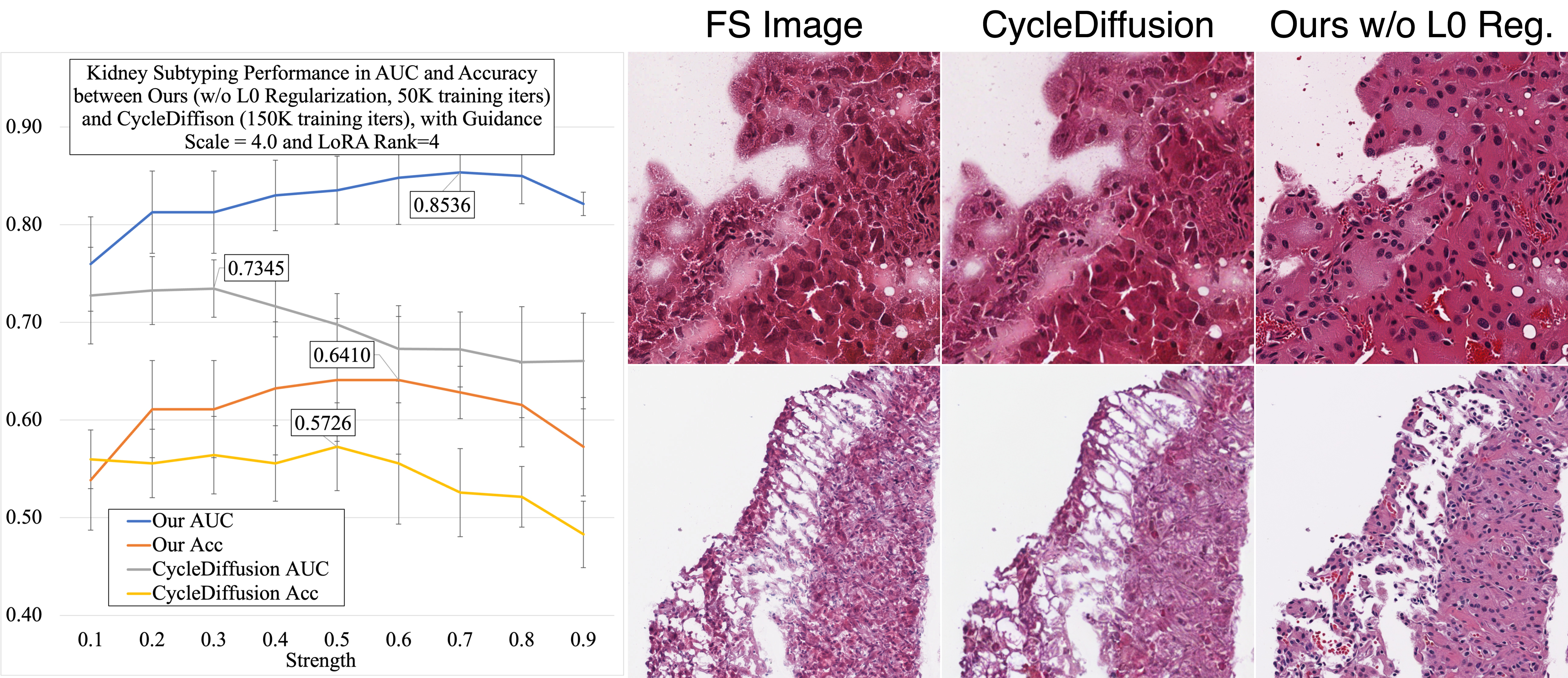}
    \caption{A comparison to CycleDiffusion with the same hyper-parameters.}
    \label{fig:supp_cd_comp}
\end{figure}

\section{Conclusion}
We benchmarked the latest Generative Adversarial Networks (GANs) and Latent Diffusion Models (LDMs) to tackle the issues of translating Frozen Section (FS) images to Formalin-Fixed Paraffin-Embedded (FFPE) images and restoring FS artifacts. To address the remaining issues, we introduce an innovative framework utilizes LDMs, enhanced with histopathology pre-trained embeddings and a FS to FFPE embedding translation mechanism, to deliver high-quality image translations that preserve essential histological details for precise clinical assessments. Our model surpasses state-of-the-art methods like AIFFPE \cite{aiffpe}, UVCGAN2 \cite{torbunov2023uvcgan2}, and CycleDiffusion \cite{wu2022cyclediffusion}, establishing new standards in FS to FFPE image translation and artifact restoration.

\bibliographystyle{splncs04}
\bibliography{refs}




\end{document}